\documentstyle [12pt,epsfig]{article}

\input epsf
\def\Bz    {{B^0_d}}
\def\Bb  {\kern 0.18em\overline{\kern -0.18em B}{}}
\def\Bzb {{\Bb^0_d}}
\def\BzBzb {B^0_d {\kern -0.16em \Bzb}}
\begin{document}

\rightline{~}
%\rightline{\today}
\bigskip
\centerline{
\LARGE \bf \boldmath  $\Bz \Bzb$ \unboldmath Mixing and \boldmath
$CPT$ \unboldmath tests}
\smallskip 
\centerline{\LARGE \bf  at Belle }
\bigskip
\bigskip
\centerline{\bf Talk given at}
\smallskip
\centerline{\bf ``Frontiers in Contemporary Physics II'',}
\smallskip
\centerline{\bf Vanderbilt University, Nashville, March 2001}
\bigskip
\begin{center}

% This command is to make the footnote counter a symbol (*, \dagger, ...)
\renewcommand{\thefootnote}{\fnsymbol{footnote}}
% This command increases the counter by one, so it displays the dagger
\stepcounter{footnote}

\bigskip
{\normalsize Christos Leonidopoulos \footnote{e-mail address: {\tt 
christos@fnal.gov}} \\ 
 {\it Columbia University, New York, New York 10027}}
\bigskip

for the Belle Collaboration
 \\[2cm]
\end{center}

% This command is to make the footnote counter an arabic number again
\renewcommand{\thefootnote}{\arabic{footnote}}
% This command is to set the footnote counter to ``1''
\setcounter{footnote}{0}

\begin{abstract} 
We present results on the mixing parameter $\Delta m_d$ and tests of
the $CPT$ symmetry in the $\Bz \Bzb$ mixing from the time evolution of 
dilepton events on the $\Upsilon(4S)$ resonance. The analysis is based 
on a 5.9 fb$^{-1}$ data sample collected by the Belle detector at
the KEKB accelerator from January to July of 2001. We obtain $\Delta
m_d$ = 0.463 $\pm$ 0.008 (stat) $\pm$ 0.016 (syst) ps$^{-1}$. No
evidence for $CPT$ violation is found. 

This is the first determination of $\Delta m_d$ from time evolution
measurements at $\Upsilon(4S)$, and the first time that an experimental
limit on $|m_{\Bz} - m_{\Bzb}\,|\, / m_{\Bz}$ has been obtained.
\end{abstract}

\newpage

\section{Introduction}

The mixing parameter of the $\Bz$-$\Bzb$ system is equal to the mass
difference of the physical (mass) eigenstates of the $\Bz$ mesons, $\Delta
m_d$. $\Delta m_d$ is considered a fundamental parameter of the
$\Bz$ system and a crucial element in the study of $CP$ violation. Its
extremely small value (13 orders of magnitude smaller than the $\Bz$
mass) enhances the sensitivity of the $\Bz ~ \Bzb$ interferometry,
making the neutral meson system a powerful probe for physics beyond
the Standard Model.

At KEKB, 8.0 GeV electrons and 3.5 GeV positrons collide on the
$\Upsilon(4S)$ resonance, producing correlated $\Bz$-$\Bzb$ pairs that
live in an antisymmetric state. The time-dependent asymmetry between
same-flavor ($\Bz \Bz \,$/$\Bzb \Bzb \,$) and opposite-flavor ($\Bz
\Bzb \,$) decay pairs undergoes an oscillation with frequency equal to
the mixing parameter, $\Delta m_d$. The center of mass (CM) of the
meson system is moving along the electron beam direction with a
Lorentz boost of $\beta \gamma = 0.452$. Since the two $B$ mesons are
practically still at the $\Upsilon(4S)$ frame, we can use this boost
and their decay vertex separation, $\Delta z$, to calculate the proper 
time difference between their decay times: $\Delta t = \Delta z /
\beta \gamma c$. The typical decay vertex separation is 200 $\mu$m, a
distance well within Belle's vertexing capabilities \cite{BELLE}.

High momentum electrons (or positrons) and muons are used for both the 
$B$ flavor tagging and the decay vertex determination. The proper
time distributions for same- and opposite-flavor $B$ decay pairs can
be used for the study of the $\Bz - \Bzb$ time evolution and the
$\Delta m_d$ measurement. By allowing for a possible non-zero mass or
lifetime difference between $\Bz$ and $\Bzb$ --- encoded in the
complex parameter $\cos \theta$ of the system's ``effective'' Hamiltonian
\cite{cpt_theory} --- we can use the same time distributions to test
the conservation of the $CPT$ symmetry in the $\Bz - \Bzb$ mixing.

The analysis presented here is based on data samples collected by the
Belle detector on the $\Upsilon(4S)$ resonance and 60 MeV below the
peak from January to July of 2001, corresponding to integrated
luminosities of 5.9 fb$^{-1}$ and 0.6 fb$^{-1}$, respectively.

\section{The Belle Detector}

The Belle detector is designed to make precise measurements of charged
and neutral particles over a wide acceptance. The combination of a
silicon vertex detector (SVD) and a central drift chamber (CDC)
surrounding the SVD is used for vertexing. For measurements of the
charged particle momenta, the SVD and CDC are embedded in a
3.4 m diameter superconducting solenoid. The generated 1.5 T magnetic
field runs parallel to the direction of the $z$-axis (defined by the
electron beam direction). 

For the particle identification (PID), three subsystems are employed:
the CDC (measuring the mean energy loss $dE/dx$), the Aerogel \v
Cerenkov Counters or ACC  (a set of 1188 \v Cerenkov radiation
modules located outside the CDC volume) and the Time of Flight system or TOF
(128 scintillation counters providing timing information).

A pair of calorimeters, the Electromagnetic Calorimeter or ECL (8736
CsI/Tl crystals) and the Electromagnetic Forward Calorimeter or EFC (320
BGO crystals), is used for the detection and identification of photons
and electrons. 

Finally, a set of 14 layers of resistive plate counters sandwiched
between 4.7 cm thick iron plates and filled up with a gas mixture on
high voltage is the $K_L$ and muon subdetector (KLM). KLM is the only
detector system placed outside the coil. The iron also serves as the
return path for the magnetic flux.

Ref.~\cite{BELLE} gives details on the specifications of the detector
components. 

\section{Analysis}

In this analysis we exclusively use leptons for flavor tagging. This mode is
referred in the bibliography as ``dileptons'': events where both $B$
mesons are flavor-tagged by leptons. The term ``dileptons'' implies
that both of the $B$  mesons decay via a semileptonic process $\Bz(t),
~\Bzb(t) ~\longrightarrow ~\ell^\pm X^\mp$ (where $\ell^\pm = e^\pm,
\mu^\pm$) through a $b \rightarrow c$ quark transition, emitting {\it
primary} leptons.

\subsection{Selection criteria}
\label{Cuts}

A set of kinematic and quality criteria is used to separate hadronic
events from continuum, QED and beam backgrounds, and to suppress
contamination from cascade or fake leptons. 

Hadronic events are classified as having at least five tracks, a primary 
vertex with radial ($V_r$) and $z$ ($V_z$) components with respect to
the origin satisfying $V_r < 1.5$ cm and $|V_z| <$ 3.5 cm, a total
visible energy detected by ECL and CDC greater than 50\% of
$E_{\Upsilon(4S)}^{\rm CM}$ ($E_{\Upsilon(4S)}^{\rm CM}$ being the
energy of $\Upsilon(4S)$ at CM), a $z$ component of the total visible
energy less than 30\% of $E_{\Upsilon(4S)}^{\rm CM}$, a total
calorimeter energy between 2.5\% and 90\% of $E_{\Upsilon(4S)}^{\rm
CM}$, and a ratio $R_2$ of second and zeroth Fox-Wolfram moments
\cite{r2} less than 0.7.

We select the electrons and the muons from the data sample of hadronic
events. For the electron identification, we use the cluster energy,
and position and shape of the electromagnetic shower in ECL, the $dE/dx$
energy loss rate in CDC, the matching between the track in the CDC and
the cluster  in ECL, and TOF and ACC hit information for the fake lepton
suppression. Optimization of the above parameters yields a $\sim$ 90\% 
efficiency for electrons and a $\sim$ 0.3\% misidentification
probability for charged tracks with $p > 1$ GeV/$c$. Electrons
consistent with pair production from photon conversions are
removed. For the muon identification, we extrapolate the track in the
CDC to the KLM, and we examine the difference between the expected and 
the actual penetration, as well as the matching quality of the KLM
hits with the fitted track. The efficiency is $\sim85\%$ for muon
tracks and the misidentification probability $\sim 2\%$ for particles
with $p > 1$ GeV/$c$. 

A ``$J/\psi$ veto'' is then applied, eliminating events consistent
with a $J/\psi$ decay to a dilepton pair. All remaining leptons of the
event are ordered according to the magnitude of their momentum at the
$\Upsilon(4S)$ rest frame. Dilepton event candidates must have at
least two particles identified as leptons. The leptons with the
highest and the second highest momenta in the CM are selected as
``dileptons''. Leptons are required to have a CM momentum and a
laboratory polar angle within the ranges $1.1 < p^{*}
 < 2.3~$GeV/$c$ and $30^\circ < \theta^{\rm lab} < 135^\circ$, $r$ and
$z$ distances of the decay vertex from the beam interaction point (BIP)
satisfying $|\Delta r\,^{\rm BIP}| < 0.05~{\rm cm}$ and ${|\Delta
z\,^{\rm BIP}| < 2.00~{\rm cm}}$,  and at least two and one hits
in the $z$ and $(r,~\phi)$ SVD layer strip planes, respectively. This is a
combination of cuts aiming at suppressing cascade and fake leptons, as
well as selecting events with  good vertex resolution for the lepton
tracks. To further reduce particle pairs originating from the same
$B$, the opening angle between the lepton tracks calculated at CM,
$\theta^*_{\ell \ell}$, is required to satisfy $-0.80 <
\cos\theta^*_{\ell \ell} < 0.95$. The above values of the cuts have
been optimized based on Monte Carlo (MC) studies. Application  of all the
selection criteria yields 8573 same-sign (SS) and 40981 opposite-sign
(OS) dilepton events on the $\Upsilon(4S)$ peak, and 40 SS and 198 OS
dilepton events below the resonance.

\subsection{Vertexing}

The proper time difference can be calculated from the $z$ component of the
distance between the decay vertices of the $B$ meson pair, $\Delta
z$: $\Delta t = \Delta z / \beta \gamma c$. The $z$-vertex position of
the leptons is determined by finding the intersection of each of the fitted
lepton tracks with an ellipsoid known to contain the $\Bz$ decay
vertices. The dimensions of this volume are calculated by convolving
the BIP ``profile'' ($\sigma_x^{\rm BIP} \sim 100 - 120~\mu{\rm m}$,
$\sigma_y^{\rm BIP} \sim 5~\mu{\rm m}$, $\sigma_z^{\rm BIP} \sim 2000
- 3000~\mu{\rm m}$) with the average  flight length of the $B$ meson
($\sim 20~\mu$m at CM). The center and the dimensions of the BIP
profile are determined by the primary vertex position distribution on
a run-by-run basis (a few tens of thousand events). The large uncertainty
of the $z$ component of the BIP is irrelevant since we are only
interested in the difference between the $z$ vertices, $\Delta
z$. $\Delta z$ is defined as $z_{\ell^+} - z_{\ell^-}$ for the OS
events, whereas for the experimentally indistinguishable leptons of SS 
events we use the absolute value of $\Delta z$.

\subsection{General strategy}

We classify the contributions to the SS and OS proper time
distributions as ``signal'' (events with two primary leptons from
semileptonic $B^\pm$ or $\Bz$ decays), ``background'' ($\Upsilon(4S)$
events with at least one cascade or fake lepton) and continuum (all
non-$\Upsilon(4S)$ events). The background is further divided into two
sub-categories indicating correctly or wrongly tagged events. The
$\Delta z$ spectra for SS and OS dileptons can therefore be described
as sums of signal ($S$), correctly tagged background ($C$), wrongly tagged 
background ($W$) and continuum. Each of these contributions is
characterized by a (normalized) proper time distribution, and is
scaled by the number of source events ($N$) and an overall selection
efficiency ($\epsilon$). Parameters that control the populations of source
events are the number of $\Upsilon(4S)$ and continuum events in the
data sample ($N_{\Upsilon(4S)} = N_{b \bar{b}}$ and $N_{\rm cont}$),
the branching fractions of neutral and charged $B$ pairs in
$\Upsilon(4S)$ decays ($f_0$ and $f_\pm=1-f_0$), the fraction of mixed 
events in neutral $\Bz$ pairs ($\chi_d$), and the semileptonic
branching fractions for neutral and charged $B$ mesons ($b_0$ and
$b_\pm$). Table~\ref{t_ssos} summarizes the categories of the events for the
SS and OS distributions and the numbers of source events, $N$.

\begin{table}
\caption{Classification of different terms in SS and OS
dilepton samples and numbers of contributing events. }
\label{t_ssos}
\begin{center}
\begin{tabular}{|c|c|c|c|c|} \hline \hline
{\bf Source} & \boldmath $N$ \unboldmath
& {\bf Event Tag} & {\bf (SS) } & {\bf (OS) } \\ \hline  \hline
\boldmath$\Bz\Bz$\unboldmath & 
$N_{\Upsilon(4S)} \, f_0 \, \chi_d \, b_0^2$ & $S$ & $\surd$ & --- 
\\ \cline{2-5}
{\bf and} & {} & $C$ &  $\surd$ & ---
\\ \cline{3-5} 
\boldmath$\Bzb\Bzb$ \unboldmath &
\raisebox{1.5ex}[0cm][0cm]{$N_{\Upsilon(4S)}\, f_0 \, \chi_d$} & $W$ &
--- & $\surd$ \\ \hline  
{} & $ N_{\Upsilon(4S)} \,f_0 \,(1-\chi_d) \,b_0^2$ & $S$ & --- &
$\surd$ \\ \cline{2-5}
\boldmath$\Bz\Bzb$\unboldmath & {} & $C$ & --- &  $\surd$\\ \cline{3-5} 
{} & \raisebox{1.5ex}[0cm][0cm]{$N_{\Upsilon(4S)} \,f_0 \,(1-\chi_d)$}
& $W$ &  $\surd$ & --- \\ \hline 
{} & $ N_{\Upsilon(4S)} \,f_\pm  \,b_\pm^2$ & $S$ & --- &
$\surd$ \\ \cline{2-5}
\boldmath$B^+ B^-$\unboldmath & {} & $C$ & --- &  $\surd$\\ \cline{3-5} 
{} & \raisebox{1.5ex}[0cm][0cm]{$N_{\Upsilon(4S)} \,f_\pm $}
& $W$ &  $\surd$ & --- \\ \hline 
{\bf Continuum} & $N_{\rm cont}$ & --- & $\surd$ &$\surd$  \\ \hline
\hline
\end{tabular}
\end{center}
\end{table}

The signal terms for mixed ($\Bz\Bz \,/ \Bzb\Bzb$: ``mix''), unmixed
($\Bz\Bzb$: ``unm'') and charged ($B^+ B^-$: ``chd'') $B$ meson pairs are
modeled analytically by smearing theoretical expressions with explicit
mixing and $CPT$-violating dependence: 

\begin{eqnarray} 
 P_{\rm mix}(\Delta t) &=& (|\sin\theta|^2 / 4\tau_{\Bz} ) 
   \,e^{-|\Delta t|/\tau_{\Bz}} \left[1 - \cos(\Delta m \, \Delta t) \right]\\
 P_{\rm unm}(\Delta t) &=& (1 / 4\tau_{\Bz} ) 
      \,e^{-|\Delta t|/\tau_{\Bz}} [1 + |\cos\theta|^2
%\nonumber \\ & &
      + (1 - |\cos\theta|^2)\cos(\Delta m \, \Delta t) 
\nonumber \\ & &
     \hspace{5.0cm} - 2\,Im(\cos\theta)\sin(\Delta m \, \Delta t) ]  \\
P_{\rm chd}( \Delta t) &=& (1 / 2\tau_{B^\pm}) \, e^{-|\Delta
t|/\tau_{B^\pm}} 
\label{Imcos}
\end{eqnarray}
A non-zero value for the complex parameter $\cos\theta$
would indicate $CPT$ violation \cite{thesis}. A $\cos \theta \ne 0$
would also affect the $\Bz$ mixed event fraction, $\chi_d \,$: 
\begin{equation}
 \chi_d = {{ |\sin\theta|^2 \,x_d^2 } \over
          { |\sin\theta|^2 \,x_d^2 + 
            (2 + x_d^2 + x_d^2 |\cos\theta|^2) }  }
\label{chi0}
\end{equation}
where $x_d \equiv \tau_{\Bz} \,\Delta m_d \,$; $1/\tau_{\Bz} \equiv
(\Gamma_H + \Gamma_L)/2$ is the mean width of the mass eigenstates,
assumed to be known in this analysis \cite{PDG}.

If $CPT$ is a good symmetry, then $\cos \theta = 0$, and the above
expressions become the standard ones one finds in the bibliography:

\begin{eqnarray} 
 P_{\rm mix}^\prime(\Delta t) &=& (1 / 4\tau_{\Bz} ) 
   \,e^{-|\Delta t|/\tau_{\Bz}} \left[1 - \cos(\Delta m \, \Delta t) \right]\\
 P_{\rm unm}^\prime(\Delta t) &=& (1 / 4\tau_{\Bz} ) 
      \,e^{-|\Delta t|/\tau_{\Bz}} \left[1 + \cos(\Delta m \, \Delta t) \right] 
\end{eqnarray}
\begin{equation}
 \chi_d^\prime = {{ x_d^2 } \over {  2 (1+ x_d^2) }  }
\end{equation}
The detector resolution for the signal terms is determined by the
$\Delta z$ distribution of $J/\psi$ decays to lepton pairs, which here 
serves as a ``response function''. These leptons are selected with the same
kinematic and quality criteria applied on primary lepton pairs from
$B$ decays\footnote{With the exception of the $J/\psi$ veto, of
course.} (Sec.~\ref{Cuts}). The $\Delta z$ distribution is
obtained by finding two separate $z$ vertices for the two lepton
tracks of the $J/\psi$ decay, and by taking their relative
difference. Its width is $\sigma=112~\mu$m. To properly account for
the long tails at large $|\Delta z|$, the distribution is binned in a
lookup table $g$, which is used to smear the theoretical functions: 

\begin{equation}
\tilde P_a(\Delta t_{\rm smear})~=~{{\int g(\Delta t_{\rm smear}-\Delta t)
P_a(\Delta t) \,d(\Delta t)} \over {\int \int g(\Delta
t_{\rm smear}-\Delta t) P_a(\Delta t) \,d(\Delta t)\, d(\Delta t_{\rm smear})}}
\label{smearfit}
\end{equation}
with $a$ = ``mix'', ``unm'', ``chd''. 

The modeling of the $\Upsilon(4S)$ background distributions is done
numerically. To this end, we have simulated large MC event samples of
generic charged and neutral $B$ meson $\Delta z$ distributions. The
major component is the combination of a primary and a secondary lepton 
from a cascade decay ($c \rightarrow s$ or $\tau$ decay). We separate
the contributions from cascade and fake leptons. The number of cascade
leptons is scaled to match the ${\cal B}(B \longrightarrow D^\pm~X)$, 
${\cal B}(B \longrightarrow D^0~X)$ branching ratios measured by
CLEO~\cite{cleo1}. The number of fake leptons is independently
determined with $K_S \rightarrow \pi^+ \pi^-$ decays from the same run
period. The small discrepancy in the detector's vertexing performance found between
MC and data ($\sim$10\%) is corrected by applying a convolution with a 
single Gaussian with $\sigma=50^{+18}_{-12}~\mu$m to each MC-determined
background distribution. This procedure is tested on similarly
obtained $\Delta z$ distributions from $J/\psi \rightarrow \ell^+
\ell^-$ and $K_S \rightarrow \pi^+ \pi^-$ decays and is found to give
a satisfactory matching. The strong dependence of the background
distributions on the value of $\Delta m_d$ is modeled through a
linear interpolation of distributions from two generic $\Bz$ MC
samples, generated with $\Delta m_d=0.423~$ps$^{-1}$ and $\Delta
m_d=0.464~$ps$^{-1}$. Finally, we have the option of adjusting the
steepness of the exponentially falling background distributions. This
is useful when we want to perform the fit for a different value of the
(fixed) $\Bz$, $B^\pm$ lifetimes, or calculate the associated
systematic error. 

The continuum $\Delta z$ distributions are also modeled numerically
with MC simulation samples. The relative weight of the continuum
contribution is determined by the number of off-resonance data, scaled
to account for luminosity and energy differences. The correction to
the vertex resolution is applied here as well. 

\subsection{Fitting}

The data is fit to a linear combination of analytical and numerical
expressions to extract the mixing and the $CPT$ violation
parameters. A binned maximum likelihood fit is performed
simultaneously to the SS and OS dilepton $\Delta z$ distributions in
the region $|\Delta z| <$ 1.85~mm \cite{thesis}.  

We fix the parameters $\tau_{\Bz}=1.548$ ps \cite{PDG},
$\tau_{B^\pm}/\tau_{\Bz} = 1.06$ \cite{PDG}, $f_\pm/f_0 = 1.05$
\cite{cleo2} and $b_\pm/b_0 = \tau_{B\pm}/\tau_{\Bz}$
\cite{thesis}. Besides $\Delta m_d$ and the (complex) parameter $\cos
\theta$, the fit includes two more free parameters, the selection
efficiency ratios $\epsilon_S / \epsilon_C$ and $\epsilon_S /
\epsilon_C$ for the charged $B$ mesons. The relative selection
efficiencies for mixed, unmixed and charged $B$  mesons within each
event tag type ($S$, $C$ or $W$) are fixed to their MC values. 

The expected total MC distributions for the SS, OS spectra are not
normalized separately to the numbers of the events of the two
histograms, but only to the total number of data events $N_{\rm tot} =
N_{SS}+N_{OS}$. In other words, the ratio of the populations of the
SS and OS total MC distributions is determined by the fit result for the mixing.

Assuming that $CPT$ is a good symmetry, and therefore by fixing $\cos \theta = 0$, 
we find $\Delta m_d = 0.463 \pm 0.008~{\rm ps}^{-1}$ with
$\chi^2/N_{\rm dof} = 332.5/376$. The efficiency ratios are
$\epsilon_S / \epsilon_C = 143.2 \pm 5.1$ and $\epsilon_S / \epsilon_W
= 103.6 \pm 15.0$.

When $\cos \theta$ is a free parameter in the fit, we find $\Delta m_d
= 0.461 \pm 0.008~{\rm ps}^{-1}$, $Re(\cos\theta) = 0.00 \pm 0.15$ and
$Im(\cos\theta) = 0.035 \pm 0.029$, with $\chi^2/N_{\rm dof} =
331.0/376$. These results are consistent with $CPT$ symmetry. The
efficiency ratios in this case are $\epsilon_S / \epsilon_C = 142.8
\pm 5.1$ and $\epsilon_S / \epsilon_W  = 102.8 \pm 14.9$. The
fractions of signal events are found to be 32.1\% in SS (mixed $B$
pairs) and 77.5\% in OS (32.1\% unmixed and 45.9\% charged $B$ pairs).

\begin{figure}[hbtp]
\centerline{\epsfxsize 5.3 truein
\epsfbox{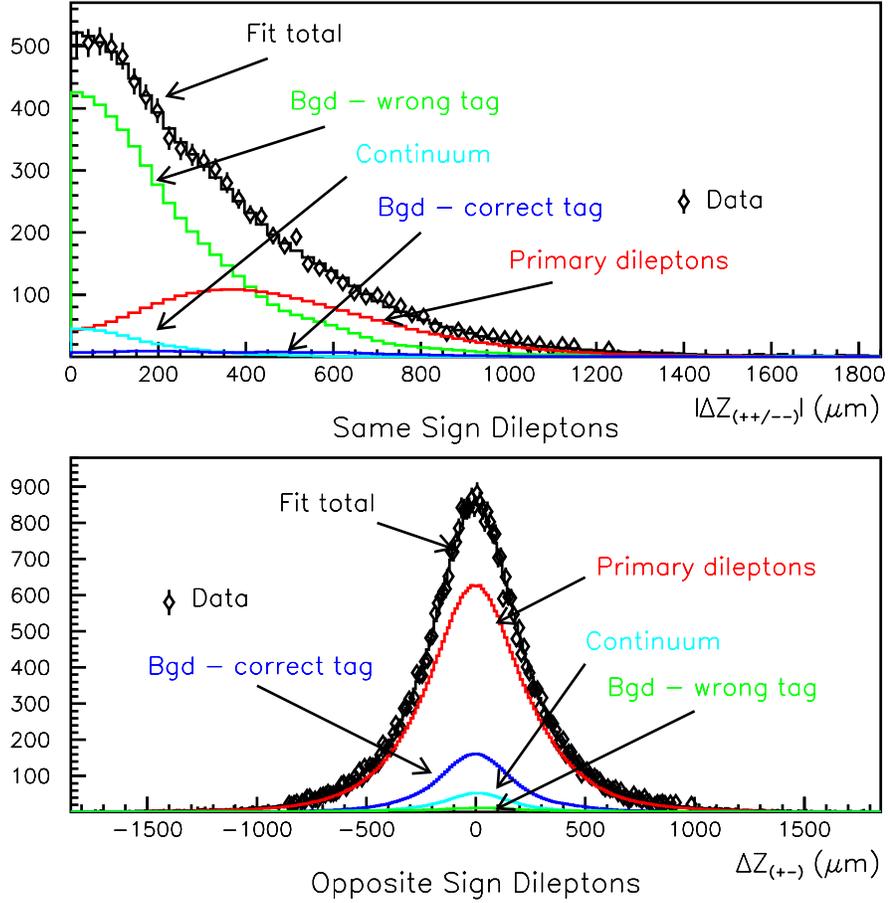}}
\caption{\label{debug} $\Delta z$ distributions for same-sign
(top) and opposite-sign (bottom) dileptons for data (points) and the fitted
Monte Carlo distributions (histograms). For the Monte Carlo distributions, signal
and background from different categories are plotted separately.} 
\end{figure}

Fig.~\ref{debug} shows the $\Delta z$ distributions for the
data and the fitted Monte Carlo distributions, for the SS (top) and OS
(bottom) dilepton spectra. The signal in the SS histogram is a small term
compared to the background from $\Bz\Bzb$ and $B^+B^-$. A much more
evocative way of displaying the $\Bz-\Bzb$ oscillation is by plotting
the time dependent asymmetry between SS and OS dileptons: 

$$A(\Delta t) = {{N^{+-}(\Delta t) - N^{\pm \pm}(\Delta t)}  \over
{N^{+-}(\Delta t) + N^{\pm \pm}(\Delta t)}} \sim \cos (\Delta m \,
\Delta t)$$

\begin{figure}[hbtp]
\centerline{\epsfxsize 5.3 truein
\epsfbox{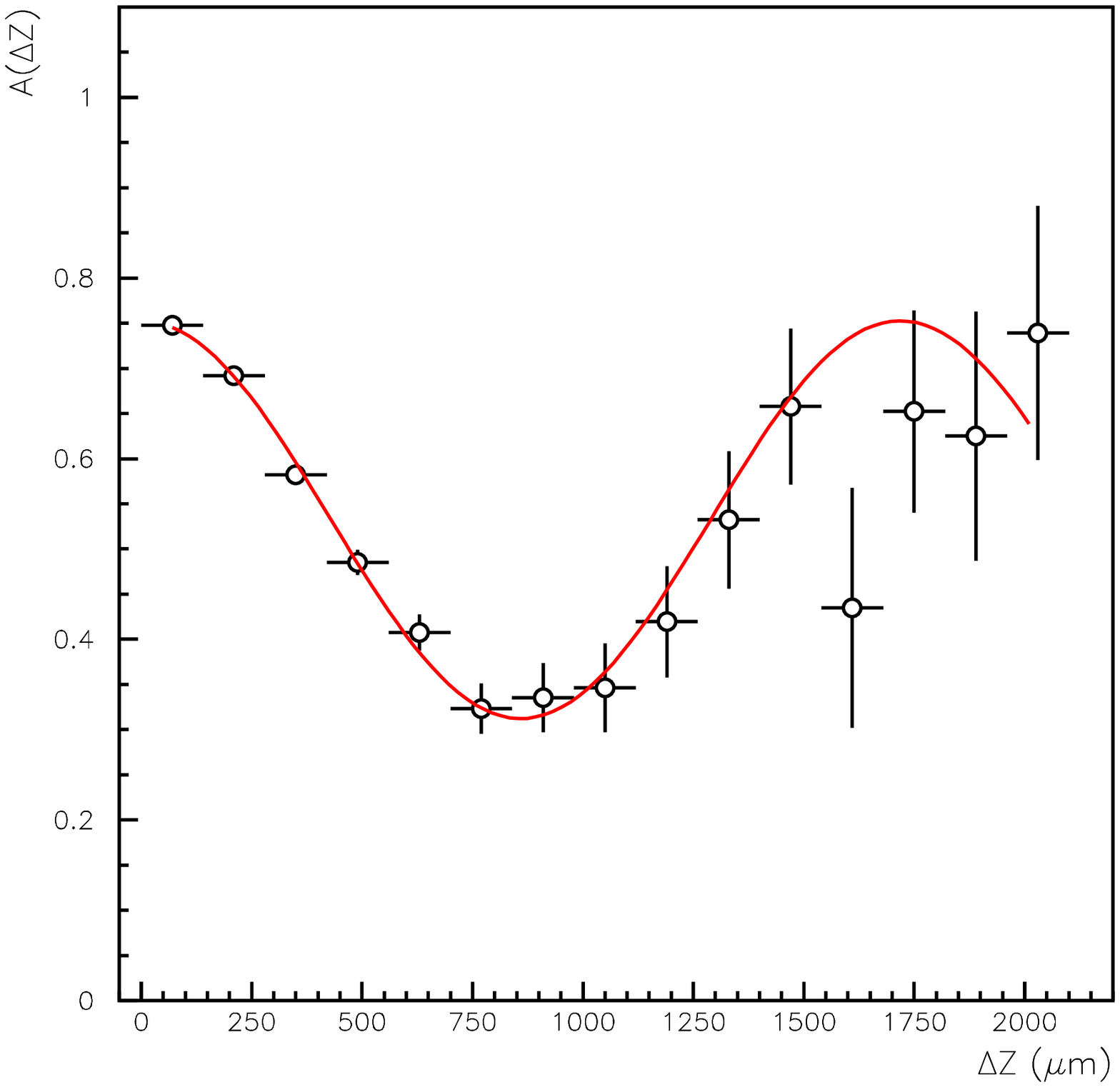}}
\caption{\label{asymmetry} Time dependent asymmetry between SS and OS
dileptons. The curve is a convoluted cosine with period determined by
the fit. The negative $\Delta z$ region for the OS histogram has been
folded into the positive region for display purposes.}
\end{figure}

To a good approximation, most of the background cancels out, and the
time dependent asymmetry reveals a clean oscillation signal, as
illustrated in Fig.~\ref{asymmetry}. In the same plot, a convolved
cosine term is superimposed on the data points, with the period
determined from the fit.

\subsection{Systematic errors}

To estimate the systematic errors of the various assumptions made in
the fit, the associated input parameter is varied, typically by $\pm
\sigma$, and then the fit is repeated. The shift in the central value
is taken to be the systematic error for that category. 

The main contributions to the systematic errors are from the
uncertainty in the measurements of the $B$ lifetimes and the branching
fractions of $\Upsilon(4S)$ to charged and neutral $B$ pairs. We
calculate these contributions by adjusting the corresponding parameters by 
the amount of their uncertainties, one at a time. The lifetime changes 
affect all $B$-originating terms in the fit (exponentially decaying
distributions) and, indirectly, signal or background events with one
primary lepton through the propagated change in the semileptonic
branching ratios. The change of branching fractions $f_\pm$, $f_0$
affects the relative populations of charged and neutral $B$ pairs in
the distributions (signal and background).

We have made the approximation that the difference between the decay widths of the
$\Bz$ mass eigenstates, $\Delta \Gamma_d$, is zero. The conservative
estimation used by most phenomenologists is ${\Delta \Gamma_d /
\Gamma_d} \leq 10^{-2}$ \cite{PDG}. We report the shift in the fitted values by
assuming a ${\Delta \Gamma_d / \Gamma_d} = 1\%$ and by modifying
accordingly the proper time distributions~\cite{thesis}. For the
systematic errors associated with the response function, we estimate
the effect of the $B$ motion in the $\Upsilon(4S)$ frame (which
includes the discrepancy of the $J/\psi$ $\Delta z$ distribution from
the true dilepton response function) and the error originating from
the statistical uncertainties in the determination of the response
function. This set of changes only affects the signal terms.

We change the cascade lepton rates by modifing the branching ratios of
$B$ mesons decaying to $D^0$ and $D^\pm$ in the MC by $\pm4.6\%$ and $\pm14.3\%$,
respectively. We change the levels of fake lepton and continuum contamination in the
sample by shifting the fake rates by $\pm$35\%, and the continuum
contribution by $\pm16$ (SS) and  $\pm7\%$ (OS). The error in the
determination of the width of the convolving Gaussian ($\sigma =
50^{+12}_{-18}~\mu$m) is used as the uncertainty in the detector's
vertexing performance for the background terms ($\sim 30\%$ of the sample). We
change the width of the Gaussian and report the shift in the fit
central values with the new background distributions. We check the
effect of finite MC statistics by varying the number of background
events on a bin-by-bin basis by the amount of the statistical errors. 

We finally check the effect of binning and several other possible
sources of systematic errors, and we find that their contribution is
negligible.

Table~\ref{sys_t} summarizes the systematic errors for $\Delta m$,
$Re(\cos\theta)$ and $Im(\cos\theta)$. Since the asymmetry between
positive and negative errors is small ($\sim 15-20\%$), we
conservatively take the larger of the values to be the systematic error.

\begin{table}
\caption{Summary of systematic errors (a) with and (b) without
invoking the $CPT$ symmetry.}
\label{sys_t}
\begin{center}
\begin{tabular}{|c||c||c|c|c|} \hline \hline 
Input parameter & \multicolumn{4}{|c|}{\bf Fitting for} \\ \cline{2-5}
and &  {\bf (a)~\boldmath $\Delta m_d$ \unboldmath} &  
\multicolumn{3}{|c|}{\bf (b)~ \boldmath $\Delta m_d$ and $\cos \theta$
\unboldmath}  \\ \cline{2-5}
uncertainty & $\Delta m_d$ (ps$^{-1}$) &  $\Delta m_d$ (ps$^{-1}$) &
$Re(\cos\theta)$ & $Im(\cos\theta)$ \\ \hline \hline
\multicolumn{5}{|c|}{$B$ lifetimes and semileptonic branching ratios}
\\ \hline
& $-0.004$ & & & $+0.004$ \\ 
\raisebox{1.5ex}[0cm][0cm]{$\tau_{B^0}=1.548 \pm 0.032$~ps} & $+0.003$ &
\raisebox{1.5ex}[0cm][0cm]{$\mp 0.004$} & \raisebox{1.5ex}[0cm][0cm]{$<0.01$} &
$-0.003$\\ \hline 
& $+0.006$ & $+0.006$ & & \\ 
\raisebox{1.5ex}[0cm][0cm]{$\tau_{B^\pm}/\tau_{B^0}=1.06\pm0.03$} &
$-0.009$ & $-0.008$ & \raisebox{1.5ex}[0cm][0cm]{$<0.01$} &
\raisebox{1.5ex}[0cm][0cm]{$<0.001$} \\ \hline 
$\Delta \Gamma_d/\Gamma_d <1\%$ & $<0.001$ & $<0.001$ &$\pm0.01$ &
$<0.001$ \\ \hline  \multicolumn{5}{|c|}{Fractions of charged and
neutral $B$ meson pairs} \\ \hline 
& $+0.007$ & $+0.008$ & &  \\
\raisebox{1.5ex}[0cm][0cm]{$f_\pm / f_0=1.05\pm0.08$} &
$-0.009$  & $-0.010$ &  \raisebox{1.5ex}[0cm][0cm]{$<0.01$}  &
\raisebox{1.5ex}[0cm][0cm]{$\mp0.001$}\\ 
\hline
\multicolumn{5}{|c|}{Response function} \\ \hline
$B$ motion in $\Upsilon(4S)$ frame& $\pm0.001$ & $\pm0.001$ &
 $\pm0.06$ & $<0.001$\\ \hline  
 statistics of $J/\psi$ sample & $\pm 0.005$ & $\pm 0.005$ &$<0.01$ &
 $\pm0.050$ \\ \hline  
\multicolumn{5}{|c|}{Other background related parameters} \\ \hline
{${\cal B}(B \rightarrow D^0\,X)$ ($\pm4.6\%$)} &$<0.001$ & $<0.001$& $<0.01$ &
$<0.001$\\ \hline  
${\cal B}(B \rightarrow D^\pm\,X)$ ($\pm14.3\%$) & $<0.001$ &
$<0.001$ & $<0.01$ & $\mp0.001$ \\ \hline  
& $+0.003$& $ $ & & \\
\raisebox{1.5ex}[0cm][0cm]{Fake rates ($\pm 35\%$)} &
$-0.004$ & 
\raisebox{1.5ex}[0cm][0cm]{$\pm0.004$} & 
\raisebox{1.5ex}[0cm][0cm]{$<0.01$} &
\raisebox{1.5ex}[0cm][0cm]{$\pm0.001$} \\ \hline  
Continuum & $+0.001$ & $+0.002$& &  \\
{(SS: $\pm 16\%$, OS: $\pm 7\%$)} 
& $-0.002$ & $-0.001$  & \raisebox{1.5ex}[0cm][0cm]{$<0.01$}  &
\raisebox{1.5ex}[0cm][0cm]{$\mp0.001$} \\ \hline 
&  & $-0.001$ & & $+0.002$ \\ 
\raisebox{1.5ex}[0cm][0cm]{Detector resolution ($^{+18}_{-12}~\mu$m)} 
& \raisebox{1.5ex}[0cm][0cm]{$\mp0.001$} & $+0.000$ 
& \raisebox{1.5ex}[0cm][0cm]{$<0.01$} 
& $-0.000$ \\ \hline 
MC statistics & $\pm0.004$ & $\pm0.005$ &$<0.01$ & $\pm0.009$ \\ \hline
\multicolumn{5}{|c|}{Other systematic errors} \\ \hline
Binning & $\pm0.001$ & $\pm0.001$ &$<0.01$ & $\pm0.002$ \\ \hline \hline
 & \boldmath$+0.012$\unboldmath 
 & \boldmath$+0.014$\unboldmath & & \\
\raisebox{1.5ex}[0cm][0cm]{\bf Total}
 & \boldmath$-0.016$\unboldmath 
 & \boldmath$-0.016$\unboldmath &
 \raisebox{1.5ex}[0cm][0cm]{\boldmath$\pm0.06$\unboldmath} & 
\raisebox{1.5ex}[0cm][0cm]{\boldmath$\pm0.051$\unboldmath} \\ \hline 
\hline
\end{tabular}
\end{center}
\end{table}

\section{Summary}

We have measured the mixing parameter $\Delta m_d$ and searched for
$CPT$ violation in the $\Bz$ system from the time evolution of
dileptons in $\Upsilon(4S)$ decays. The analysis corresponds to an
integrated luminosity of 5.9 fb$^{-1}$ of data collected with the
Belle detector. The run period was from January to July of 2000. \\
If we invoke $CPT$ symmetry for $\Delta m_d$ we obtain:
$$ \Delta m_d = 0.463 \pm 0.008~{\rm (stat)}~\pm 0.016~{\rm
(sys)}~{\rm ps}^{-1}$$
If we fit simultaneously for $\Delta m_d$ and the $CPT$-violating
parameter $\cos\theta$ we obtain:
%\nonumber \\ 
\begin{eqnarray}
\Delta m_d & = & 0.461 \pm 0.008~{\rm (stat)}~\pm0.016~{\rm 
(sys)}~{\rm ps}^{-1} \nonumber \\
Re(\cos\theta) & = & 0.00 \pm 0.15~{\rm (stat)}~{\pm0.06}~{\rm
(sys)} \nonumber \\
Im(\cos\theta) & = & 0.035 \pm0.029~{\rm
(stat)}~{\pm0.051}~{\rm (sys)} \nonumber
\end{eqnarray}
These results imply \cite{thesis} for the mass and the lifetime
differences between $\Bz$ and $\Bzb \,$:  
$${{|m_{\Bz}-m_{\Bzb}|} / m_{\Bz}}~<~ 1.6 \times 10^{-14}~{\rm
and~~}{|\Gamma_{\Bz}-\Gamma_{\Bzb}| / \Gamma_{\Bz} }~<~0.161,~{\rm at~90\%~C.L.}$$
The results are consistent with $CPT$ conservation. This is the first
direct measurement and the first published result of $\Delta m_d$ from 
a time-dependent analysis on the $\Upsilon(4S)$ resonance
\cite{prl}. The limit on $ (\Gamma_{\Bz} - \Gamma_{\Bzb})/ \Gamma_{\Bzb}$
measurement is consistent with previous measurements
\cite{dg_meas}. This is the first measurement  on $Re(\cos\theta)$
and on the mass difference between $\Bz$ and $\Bzb$.

\end{document}